\documentstyle[11pt]{article}

\title{\bf On the Nikolaev-Zakharov-Zoller form for the BFKL pomeron}
\author{M.A.Braun  \\ Department of high-energy physics,
\\ University of S. Petersburg, 198904 S. Petersburg
, Russia and\\
Department of Particle Physics, University of Santiago de Compostela,\\
15706 Santiago de Compostela, Spain
}
\def\beq{\begin{equation}}
\def\eeq{\end{equation}}
\def\noi{\noindent}

\def\bk{{\rm K}_{0}}
\begin{document}
\maketitle
\medskip
\noi{\bf Abstract.}
The equation proposed by N.N.Nikolaev, B.G.Zakharov and V.R.Zoller for the
colour
dipole cross-section is compared with the BFKL equation for the hard pomeron
for
the $SU(2)$ colour group. It is demonstrated that for a fixed coupling constant
the difference between the two equations is due to a different manner to
introduce the gluon mass.\vspace{3 cm}

{\Large\bf US-FT/19-94}

\newpage
{\bf 1. Introduction.} In a series of papers N.N.Nikolaev, B.G.Zakharov and
V.R.Zoller proposed what they call a generalized BFKL equation for the
scattering
cross-section of a colour dipole [1,2]. Their equation contains the running
coupling constant introduced on the basis of the naive dimensional reasoning:
its argument is the smallest distance involved. In this sense the
Nikolaev-Zakharov-Zoller (NZZ) equation  indeed generalizes the BFKL one, valid
for a small fixed coupling constant [ 3,4 ]. However this is not the first
generalization of this kind: the running coupling depending on some presumably
appropriate scale had been introduced in many previous papers [ 5-8 ].
The employed naive method to make the coupling run may work when the
various scales involved have all different orders of magnitude but not when
some of them are of the same order. A more constructive way to introduce the
running coupling, based on the so-called bootsrap equation, has been proposed
by the author [ 9 ].

However it is not the running coupling that we are going to discuss in the
present note.
The form of the NZZ equation is very different from the original BFKL equation
even for a fixed coupling.
One might think therefore that it presents some generalization already on this
level. This, however, cannot be true, since, after all, the NZZ equation sums
the
same diagrams as described by the BFKL one and in the same approximation. In
this
note we study this point in some detail. We demonstrate that the NZZ equation
with a fixed coupling constant is essentially nothing but the BFKL equation
transformed to different variables. The difference between the two can be
traced
to the manner to introduce the gluon mass $m$. If it is introduced by a
spontaneous breakdown of symmetry, as in the original BFKL treatment, then a
certain term  is lacking in the NZZ derivation and should be added to their
equation. In the limit
$m\rightarrow 0$ the two equations are completely equivalent, which was noted
in [ 1,2 ].

Since in the comparison of the two equations the nonzero gluon mass is
important, we restrict our study to the case of $SU(2)$ colour group,
considered
in the original BFKL derivation, where the exact form of the interaction
kernel for $m\neq 0$ is known.\vspace{0.5 cm}

{\bf 2. Relation between the dipole cross-section and the BFKL amplitude.} The
starting point of the NZZ approach is the expression for the virtual
photon-target scattering cross-section [ 10 ]
\beq
\sigma_{T,L}=\int_{0}^{1}d\alpha\int d^{2}r\,\sigma(r)W_{T,L}(\alpha,r)
\eeq
Here $W_{T,L}(\alpha,r)$ is the probability for the photon to go into a
$q\bar q$ pair at a relative transverse distance $r$ and with a part $\alpha$
of
the longitudinal momentum carried by the quark. The indeces $T$ or $L$ refer
to the transversal or longitudinal polarization of the photon. The
cross-section $\sigma(r)$ corresponds to the scattering of the colour dipole
formed by the $q\bar q$ pair off the target. In the lowest approximation it is
given by the two gluon exchange:
\beq
\sigma(r)=(16\alpha_{s}^{2}/3)\int d^{2}q\,V(q)(1-\exp (iqr))/(q^{2}+m^{2})^{2}
\eeq
where $V(q)$ is the vertex for the interaction of the gluon of transverse
momentum $q$ with the target and $m$ is "the gluon mass", introduced to cure
the
infrared divergence. The NZZ equation sums corrections to $\sigma(r)$ which
result when one goes beyond the two-gluon approximation and takes into account
all terms of the order $(\alpha_{s}\ln s)^{n}$ (the leading log approximation).
To deduce their equation NZZ study the scattering of configurations of the
projectile which additionally contain a number of gluons. We shall try to
obtain it by a simpler method, by noting that  the leading log
corrections to the two-gluon exchange are summed by the BFKL equation.

For the forward virtual gluon-target (amputated) scattering amplitude
$\psi_{j}(q)$, in the complex angular momentum $j$ representation, and for
the $SU(2)$ colour group, the vacuum channel BFKL equation reads [ 3 ]
\[
(j-1)\psi_{j}(q)=V(q)+(2g^{2}/(2\pi)^{3})\int
d^{2}k_{1}(q^{2}+m^{2})(2\psi_{j}(k_{1})-\psi_{j}(q))
/(k_{1}^{2}+m^{2})(k_{2}^{2}
+m^{2})\]\beq-(5m^{2}g^{2}/2(2\pi)^{3})
\int d^{2}k\psi_{j}(k)/(k^{2}+m^{2})^{2}
\eeq
where $k_{1}+k_{2}=q$. The amplitude $\psi(s,q)$ depending on the c.m.energy
$\sqrt{s}$ is obtained from $\psi_{j}(q)$ by the Mellin transformation.
Comparing with (2) we find that to sum the leading log corrections one has to
substitite the cross-section $\sigma(r)$ by $\sigma_{j}(r)$ where
\beq
\sigma_{j}(r)=(16\alpha_{s}^{2}/3)\int d^{2}q\,\psi_{j}(q)(1-\exp
(iqr))/(q^{2}+m^{2})^{2}
\eeq
This formula establishes the relation between the cross-section $\sigma_{j}(r)$
in the NZZ treatment and the BFKL function $\psi_{j}(q)$. Evidently the
equation
for $\sigma_{j}(r)$ can be set up from the known BFKL equation (3).

We introduce the function
\beq
\phi_{j}(q)=\psi_{j}(q)/(q^{2}+m^{2})^{2}
\eeq
According to (4) the cross-section $\sigma_{j}(r)$ is nothing but the Fourier
transform of $\phi_{j}(q)$ up to a subtraction at $r=0$ and a factor:
\beq
\sigma_{j}(r)=-(32\pi\alpha_{s}^{2}/3)(\phi(r)-\phi(r=0))\equiv
-(32\pi\alpha_{s}^{2}/3)\tilde{\phi}(r)
\eeq
So all we have to do is to rewrite Eq. (3) for the function $\phi$ and
transform it to the coordinate space.\vspace{0.5 cm}

{\bf 3. The equation for $\phi(r)$.}
Substituting $\psi$ for $\phi$, Eq. (5),  we obtain an
equation in the momentum space (we suppress the index $j$ in the following)
\[ (j-1)\phi(q)=V(q)/(q^{2}+m^{2})^{2}+(4g^{2}/(2\pi)^{3})\int
d^{2}k_{1}((k_{1}^{2}+m^{2})\phi(k_{1})/(q^{2}+m^{2})(k_{2}^{2}+m^{2})\]\[-
(2g^{2}/(2\pi)^{3})(q^{2}+m^{2})\phi(q)\int
d^{2}k_{1}/(k_{1}^{2}+m^{2})(k_{1}^{2}+m^{2})\]\beq
-(5m^{2}g^{2}/2(2\pi)^{3}(q^{2}+m^{2})^{2})
\int d^{2}k\phi(k)
\eeq
In the following the inhomogeneous term will not play any role and we drop it,
limiting ourselves with the homogeneous equation for $\phi$ (and
$\sigma_{j}(r)$). Transforming to the coordinate space we get
\beq
(j-1)\phi(r)=(4g^{2}/(2\pi)^{3})I_{1}-(2g^{2}/(2\pi)^{3})I_{2}-
(5m^{2}g^{2}/2(2\pi)^{3})I_{3}
\eeq
where  $I_{1}$, $I_{2}$ and $I_{3}$ denote the integrals of the first, second
and third terms on the righthand side of Eq. (7):
\beq
I_{1}=(1/2\pi)\int d^{2}qd^{2}k_{1}\exp(iqr)
(k_{1}^{2}+m^{2})\phi(k_{1})/(q^{2}+m^{2})(k_{2}^{2}+m^{2})
\eeq
\beq I_{2}=(1/2\pi)\int d^{2}qd^{2}k_{1}\exp(iqr)
(q^{2}+m^{2})\phi(q)/(k_{1}^{2}+m^{2})(k_{2}^{2}+m^{2})
\eeq
and
\beq I_{3}=(1/2\pi)\int d^{2}q\exp(iqr)/(q^{2}+m^{2})^{2}
\int d^{2}k\phi(k)
\eeq
Of these integrals the last one is calculated directly. Taking into account
that
\beq (1/2\pi)\int d^{2}q\exp (iqr)/(q^{2}+m^{2})={\mbox K}_{0}(mr)
\eeq
and that
\beq (1/2\pi)\int d^{2}q\exp (iqr)/(q^{2}+m^{2})^{2}=-(\partial/\partial m^{2})
{\mbox K}_{0}(mr)=(r/2m){\mbox K}_{1}(mr)
\eeq
we obtain
\beq
I_{3}=(\pi r/m){\mbox K}_{1}(mr)\phi(r=0)
\eeq\vspace{0.5 cm}

{\bf 4. Transformation to the NZZ form.}
We now transform $I_{1}$ and $I_{2}$ into the form which will lead us to the
NZZ-like equation. Take $I_{1}$ and write it as a triple integral
\beq I_{1}=(1/2\pi)\int d^{2}qd^{2}k_{1}d^{2}k_{2}
\delta^{2}(k_{1}+k_{2}-q)\exp(iqr)
(k_{1}^{2}+m^{2})\phi(k_{1})/(q^{2}+m^{2})(k_{2}^{2}+m^{2})
\eeq
Presenting the $\delta$-function as an integral over transverse coordinate
vector $r'$ we rewrite (15) as
\beq I_{1}=(1/(2\pi)^{3})\int d^{2}r'd^{2}qd^{2}k_{1}d^{2}k_{2}
\exp(ir'(k_{1}+k_{2}-q)+iqr)
(k_{1}^{2}+m^{2})\phi(k_{1})/(q^{2}+m^{2})(k_{2}^{2}+m^{2})
\eeq
With (12) and
\beq (1/2\pi)\int d^{2}q\exp (iqr)(q^{2}+m^{2})\phi(q)=(m^{2}-\Delta)\phi(r)
\eeq
we obtain
\beq
I_{1}=\int d^{2}r'\bk (mr')\bk (m|r-r'|)(m^{2}-\Delta')\phi(r')
\eeq
Symmetrizing in $r_{1}\equiv r'$ and $r_{2}=r-r'$ we finally have
\beq I_{1}=(1/2)\int d^{2}r_{1}\bk (mr_{1})\bk
(mr_{2})(m^{2}-\Delta)(\phi(r_{1})+\phi(r_{2}))
\eeq
where $r_{1}+r_{2}=r$.
The same calculation for $I_{2}$ leads to
\beq I_{2}=(1/(2\pi)^{3})\int d^{2}r'd^{2}qd^{2}k_{1}d^{2}k_{2}
\exp(ir'(k_{1}+k_{2}-q)+iqr)
(q^{2}+m^{2})\phi(q)/(k_{1}^{2}+m^{2})(k_{2}^{2}+m^{2})
\eeq
which according to (12) and (17) gives
\beq I_{2}=\int d^{2}r'\bk^{2} (mr')(m^{2}-\Delta')\phi(|r-r'|)
\eeq
After the symmetrization
\beq I_{2}=(1/2)\int d^{2}r_{1}(\bk^{2} (mr_{1})(m^{2}-\Delta)\phi(r_{2})+
\bk^{2} (mr_{2})(m^{2}-\Delta)\phi(r_{1}))
\eeq
Noting that in Eq.(8) the term $I_{1}$ has a coefficient twice larger than
$I_{2}$ we can now rewrite the equation in the form
\[
(j-1)\phi(r)=-(g^{2}/(2\pi)^{3})\int
d^{2}r_{1}(\bk(mr_{1})-\bk(mr_{2}))^{2}(m^{2}-\Delta)(\phi(r_{1})+
\phi(r_{2}))\]\[+(2g^{2}/(2\pi)^{3})\int
d^{2}r'\bk^{2}(mr')(m^{2}-\Delta)\phi(r')\]\beq-
(5m^{2}g^{2}/2(2\pi)^{3})(\pi r/m){\mbox K}_{1}(mr)\phi(r=0)
\eeq
The second term on the righthand side subtracts the redundant contributions
from the first one. It does not depend on $r$.

To pass to the cross-section $\sigma_{j}(r)$ we have to make a subtraction at
$r=0$. The first term on the righthand side of (23) vanishes at $r=0$. So
putting
$r=0$  we get a relation
\beq
(j-1)\phi(r=0)=(2g^{2}/(2\pi)^{3})\int d^{2}r'\bk^{2}(mr')(m^{2}-\Delta)
\phi(r')-
(5g^{2}\pi/2(2\pi)^{3})\phi(r=0)
\eeq
Subtracting it from Eq. (23) we obtain
\[ (j-1)\tilde{\phi}(r)=-(g^{2}/(2\pi)^{3})\int
d^{2}r_{1}(\bk(mr_{1})-\bk(mr_{2}))^{2}(m^{2}-\Delta)(\phi(r_{1})+
\phi(r_{2}))\]\beq+
(5\pi g^{2}/2(2\pi)^{3})\kappa (mr)\phi(r=0)
\eeq
where we have defined
\[\kappa(z)=1-z{\mbox K}_{1}(z)\]

The next step consists in passing the Laplace operator applied to $\phi$ onto
the
Bessel functions by integrating by parts. To be able to do it we have to
regularize the integrand at $r'=0$. To this end we change
\[(m^{2}-\Delta)(\phi(r_{1})+\phi(r_{2}))\rightarrow
(m^{2}-\Delta)(\phi(r_{1})+\phi(r_{2})-\phi(r))+m^{2}\phi(r)\]
The equation becomes
\[ (j-1)\tilde{\phi}(r)=-(g^{2}/(2\pi)^{3})\int
d^{2}r_{1}(\bk(mr_{1})-\bk(mr_{2}))^{2}(m^{2}-\Delta)(\phi(r_{1})+
\phi(r_{2})-\phi(r))\]\beq-
(g^{2}m^{2}\phi(r)/(2\pi)^{3})\int
d^{2}r_{1}(\bk(mr_{1})-\bk(mr_{2}))^{2}
+(5\pi g^{2}/2(2\pi)^{3})\kappa(mr)\phi(r=0)
\eeq
Note that
\[\int d^{2}r_{1}(\bk(mr_{1})-\bk(mr_{2}))^{2}=(2\pi/m^{2})\kappa (mr)\]
Presenting
\[\phi(r)=\tilde{\phi}(r)+\phi(r=0)\]
we then get
\[ (j-1)\tilde{\phi}(r)=-(g^{2}/(2\pi)^{3})\int
d^{2}r_{1}(\bk(mr_{1})-\bk(mr_{2}))^{2}(m^{2}-\Delta)(\tilde{\phi}(r_{1})+
\tilde{\phi}(r_{2})-\tilde{\phi}(r))\]
\beq-(2\pi g^{2}/(2\pi)^{3})\kappa(mr)\tilde{\phi}(r)
-(3\pi g^{2}/2(2\pi)^{3})\kappa(mr)\phi(r=0)
\eeq
Now we can safely apply the Laplace operator to the Bessel functions. Evidently
\[\Delta(\bk (mr_{1})-\bk (mr_{2}))^{2}=
2(\bk (mr_{1})-\bk (mr_{2}))\Delta(\bk (mr_{1})-\bk (mr_{2}))\]\[+
2(\nabla(\bk(mr_{1})-\bk (mr_{2}))^{2}\]
We use
\[\Delta\bk (mr)=m^{2}\bk (mr)-2\pi\delta^{2}(r)\]
The factor $\phi(r_{1})+\phi(r_{2})-\phi(r)$ vanishes if either $r_{1}$ or
$r_{2}$ vanish. So the $\delta$-functions give nothing. Then (27) becomes
\[ (j-1)\tilde{\phi}(r)=(g^{2}/(2\pi)^{3})\int
d^{2}r_{1}(\tilde{\phi}(r_{1})+
\tilde{\phi}(r_{2})-\tilde{\phi}(r))\]\[
(m^{2}(\bk(mr_{1})-\bk(mr_{2}))^{2}+2(\nabla(\bk(mr_{1})
-\bk(mr_{2}))^{2})\]\beq
- (2\pi g^{2}/(2\pi)^{3})\kappa(mr)\tilde{\phi}(r)
 -(3g^{2}/2(2\pi)^{3})\kappa(mr)\phi(r=0)\eeq
We finally note that
\[ \nabla\bk (mr)=-m(r/|r|){\mbox K}_{1}(mr)\]
so that the equation turns into
\[ (j-1)\tilde{\phi}(r)=\]\[
(2g^{2}m^{2}/(2\pi)^{3})\int
d^{2}r_{1}(\tilde{\phi}(r_{1})+\tilde{\phi}(r_{2})-\tilde{\phi}(r))
((r_{1}/|r_{1}|){\mbox K}_{1}(mr_{1})-(r_{2}/|r_{2}|){\mbox
K}_{1}(mr_{2}))^{2})\]
\[+(g^{2}m^{2}/(2\pi)^{3})\int
d^{2}r_{1}(\tilde{\phi}(r_{1})+\tilde{\phi}(r_{2})-2\tilde{\phi}(r))
(\bk(mr_{1})-\bk(mr_{2}))^{2} \]
 \beq
-(3g^{2}/2(2\pi)^{3})\kappa(mr)\phi(r=0)
\eeq
 To obtain a closed equation
for $\tilde{\phi}$ one has  to express $\phi(r=0)$ in terms of
$\tilde{\phi}$, using Eq.(24):
\beq
\phi(r=0)=(j-1+g^{2}/2(2\pi)^{3}))^{(-1)}
(2g^{2}/(2\pi)^{3})\int d^{2}r'\bk^{2}(mr')(m^{2}-\Delta)\tilde{\phi(r')}
\eeq
Multiplying then the equation
by
$-32\pi\alpha_{s}^{2}/3$ one obtains the final equation for the cross-section
\[ (j-1)\sigma_{j}(r)=\]\[ (2g^{2}m^{2}/(2\pi)^{3})\int
d^{2}r_{1}(\sigma_{j}(r_{1})+\sigma_{j}(r_{2})-\sigma_{j}(r))
((r_{1}/|r_{1}|){\mbox K}_{1}(mr_{1})-(r_{2}/|r_{2}|){\mbox
K}_{1}(mr_{2}))^{2})\]
\[+(g^{2}m^{2}/(2\pi)^{3})\int
d^{2}r_{1}(\sigma_{j}(r_{1})+\sigma_{j}(r_{2})-2\sigma_{j}(r))
(\bk(mr_{1})-\bk(mr_{2}))^{2} \]
 \beq -(3g^{4}/(2\pi)^{6})(j-1+g^{2}/2(2\pi)^{3}))^{(-1)}\kappa(mr)
\int d^{2}r'\bk^{2}(mr')(m^{2}-\Delta)\sigma_{j}(r')
\eeq
\vspace{0.5 cm}

{\bf 5. Comparison wth the NZZ equation.} If we retain only the first term on
the righthand side of Eq. (31) then the resulting equation exactly coincides
with the NZZ one for a fixed coupling constant $g$ and the gluon mass $m$.
The rest of the terms do not appear in the NZZ treatment. Thus
the NZZ equation and the BFKL one are
generally
different. However one easily finds that when the gluon mass $m$ goes to zero
only the first term on the righthand of Eq.(31) survives. In this limit
the NZZ equation indeed goes over into the BFKL equation as noted in [ 1,2 ].

So we conclude that the NZZ and BFKL equations with a fixed coupling constant
are
different for a nonzero gluon mass. Evidently this difference originates from
a different manner to introduce the gluon mass. In the BFKL equation it is
introduced by the spontaneous breakdown of symmetry, via the Higgs mechanism.
This method preserves the gauge invariance and allows to eliminate
ultraviolet divergencies by the standard renormalization technique. The way
the gluon mass is introduced in the NZZ approach is purely phenomenological.
Evidently for a fixed and very small coupling constant this approach cannot be
consistent with the gauge invariance. If the constant runs and becomes large
in the region where the effects of the gluon mass are appreciable, then, of
course, no conclusive arguments against such a procedure can be put forward.
\vspace{0.5 cm}

 {\bf 6. Acknowledgments.} The author  expresses his gratitude
to the General Direction of the Scientific and Technical Investigation
(DGICYT) of Spain for financial support.\vspace{0.5 cm}

{\bf References.}

\noi 1. N.N.Nikolaev, B.G.Zakharov and V.R.Zoller, Pis'ma Zh.Eks.Teor.Fis.{\bf
59} (1994)8; Phys. Lett.{\bf B328} (1994) 486; preprint ITEP-74-94 (1994)\\
\noi 2. N.N.Nikolaev and B.G.Zakharov, Julich preprint KFA-IKP(Th)-1993-17
(1993); Phys. Lett. {\bf B327} (1994)157\\
\noi 3. V.S.Fadin, E.A.Kuraev and L.N.Lipatov, Phys. Lett. {\bf 60B} (1975)
50\\
\noi 4. E.Kuraev, L.Lipatov and V.Fadin, Sov. Phys. JETP, {\bf 45}
(1977) 199\\
Ya.Balitzky and L.Lipatov, Sov. Phys. J.Nucl. Phys.,{\bf 28} (1978)822\\
\noi 5. L.V.Gribov, E.M.Levin and M.G.Ryskin, Phys. Rep. {\bf 100} (1983) 1\\
\noi 6. L.N.Lipatov, Sov. Phys. JETP, {\bf 63} (1986) 904\\
\noi 7. P.D.B. Collins and J.Kwiecinsky, Nucl. Phys. {\bf B335} (1990) 335\\
\noi 8. J.Kwiecinski, A.D.Martin, P.J.Sutton and K.Golec-Biernat, preprint
DTP/94/08\\
\noi 9. M.Braun, Santiago univ. preprints  US-FT/11-94and US-FT/17-94 (1994)\\
\noi 10. N.N.Nikolaev and B.G.Zakharov, Z.Phys. {\bf C49} (1991) 607
 \end{document}